\documentclass{bcp}
\def\LaTeX{L\kern -.36em\raise .3ex\hbox{\sc a}\kern -.15em T\kern -.1667em%
\lower .7ex\hbox{E}\kern -.125em X}

\usepackage{subeqnarray}
\usepackage{graphicx}

\begin{document}

%\keywords{These are optional}
%\mathclass{Primary 46C20; Secondary 32G81.}
\thanks{This work was supported by the Center for Complexity Studies}
\abbrevauthors{O. Biham, N.Q. Balaban, A. Loinger, A. Lipshtat and H.B. Perets}
\abbrevtitle{Simulations of simple genetic circuits}

\title{Deterministic and Stochastic Simulations \\
of Simple Genetic Circuits}

\author{Ofer Biham, Nathalie Q. Balaban and Adiel Loinger}
\address{Racah Institute of Physics, The Hebrew University, Jerusalem 91904, 
Israel}

\author{Azi Lipshtat}
\address{Department of Pharmacology and Biological Chemistry,
Mount Sinai \\ School of Medicine,
New York, NY 10029, USA
}

\author{Hagai B. Perets}
\address{Faculty of Physics,
Weizmann Institute of Science, Rehovot 76100, Israel
}

\maketitlebcp

\abstract{

We analyze three simple genetic circuits which involve transcriptional
regulation and feedback: the autorepressor, the switch and the
repressilator, that consist of one, two and three genes, respectively.
Such systems are commonly simulated using rate equations, that
account for the concentrations of the mRNAs and proteins produced
by these genes. 
Rate equations are suitable when the concentrations
of the relevant molecules in a cell are large and 
fluctuations are negligible.
However, when some of the proteins in the circuit
appear in low copy numbers, fluctuations become important
and the rate equations fail. 
In this case stochastic methods, such as 
direct numerical integration of
the master equation or
Monte Carlo simulations are required.
Here we present deterministic and stochastic simulations
of the autorepressor, the switch and the repressilator.
We show that fluctuations give rise to quantitative
and qualitative changes in the dynamics of these systems.
In particular, we demonstrate a fluctuations-induced bistability
in a variant of the genetic switch and and noisy oscillations
obtained in the repressilator circuit.
}

%-----------------------------------------------------
\section{Introduction.} 

The production of proteins in cells is regulated 
by networks of interacting genes.
Some of these genes code for transcription factors,
which are proteins that regulate the expression of
other genes by binding to specific promoter sites
on the DNA.
Some of these proteins, called repressors, perform
negative regulation, while others, called activators,
perform positive regulation.
Post-trancriptional regulation can occur by 
translational regulation, or by post-translational 
regulation such as protein-protein interaction, 
which may modify the function of these proteins. 
These networks of interacting genes 
are at the heart of all processes in cells: 
from the regulation of the cell cycle 
to the various stress responses.

It turns out that genetic networks are typically sparse networks,
namely most genes interact only with a small number of other genes. 
Also, the networks exhibit some degree of modularity, namely
one can identify recurring modules, which
are known as network motifs
\cite{Milo2002}.

Modules found in nature are often hard to study and fully control. 
To overcome these limitations, synthetic networks can be constructed 
from well-studied components. 
Examples of such components are the 
lactose, the lambda and the tetracycline systems in bacteria.
The synthetic networks are designed to 
perform desired functions, 
determined by their architecture.
They do not require the manipulation of the structure of
proteins and other regulatory elements at the molecular
level.
The genes and promoters are often inserted into plasmids
rather than on the chromosome.
Two important examples of synthetic circuits 
are the genetic toggle switch and the repressilator.

The toggle switch consists of two genes, which negatively
regulate each other expression.
The regulation is performed at the transcriptional level,
namely each gene codes for a repressor protein that binds
to the promoter of the other gene.
Such system may exhibit bistability, namely two stable
states, where in each state one of the proteins is dominant
and the other is suppressed.
A synthetic toggle switch
was constructed in 
{\it E. coli} 
and the conditions for bistability
were examined
\cite{Gardner2000}.
The switching between its two states was demonstrated using 
chemical and thermal induction.
More recently, such circuit was found to exist in a natural
system in which two mutual repressors regulate
the differentiation of myeloid progenitors
into either macrophages or neutrophils
\cite{Laslo2006}.

The synthetic repressilator circuit, constructed in
{\it E. coli}, 
was designed to exhibit
oscillations, reminiscent of natural genetic
oscillators such as the circadian rhythms
\cite{Elowitz2000}. 
The repressilator circuit was encoded on a plasmid
that appears in a low copy number. 
The protein concentrations
were measured vs. time in single cells.
Oscillations with a period of about 150 minutes
were found, 
namely extending over several division cycles.
The oscillations were found to be 
noisy, typically maintaining phase coherence for 
times of the order of a single oscillation
period.

The dynamics of genetic networks is often simulated
using rate equation models.
These are sets of coupled ordinary differential equations,
which account for the concentrations of the mRNAs and proteins
in the network.
In general, rate equations account for average concentrations
and ignore fluctuations. 
They are suitable for systems in which the concentration of
interacting molecules are large and fluctuations are negligible.
However, proteins in cells often appear in low copy numbers and
may exhibit large fluctuations. 
Moreover, in case of transcriptional regulation, the expression
of the regulated gene may be controlled by a single protein
that binds to its promoter.
This extends the notion of low copy numbers and the
resulting fluctuations even to the case
when there is a large number of free repressors of a certain
type, since only one of them may bind to the promoter at any
given time.
Recent advances measurement techniques made
it possible to measure the fluctuations in 
copy number of proteins in single cells
\cite{Elowitz2002,Paulsson2004,Swain2002}. 
Measurements of protein levels in single cells revealed distributions
that depend on the topology of the regulatory 
network controlling the particular protein
\cite{Ozbudak2004}. 

To account for the fluctuations, simulations of genetic networks
should be done using stochastic methods such as the master
equation 
\cite{Kepler2001,McAdams1997,McAdams1999,Paulsson2000,Paulsson2000b,Paulsson2002,Paulsson2004}
or Monte Carlo simulations
\cite{Gibson2000,Gillespie1977,McAdams1997}.
The master equation provides the probability distribution of 
the concentrations of proteins in a population of cells.
From this distribution one can calculate the average concentrations
as well as the correlations and the rates of all processes.
Monte Carlo simulations enable to follow the temporal variations
in the level of gene
expression and in the concentrations of proteins in a single cell.
These results enable to extract the noise level and temporal correlation 
functions. 
They can also be used to characterize the dynamics of the system
and determine whether it exhibits a single steady state, multi-stability
or oscillatory behavior.

\begin{figure}
\includegraphics[width=10cm]{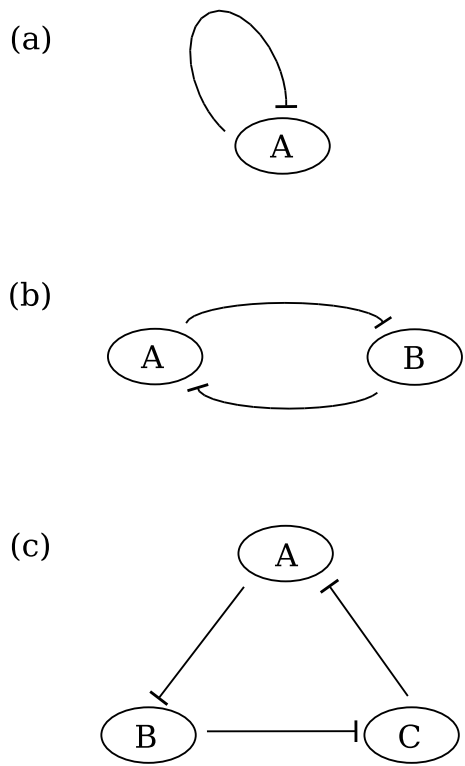}
\centering
\caption{
The three genetic circuits considered in this paper:
the autorepressor (a); the genetic switch (b); and the repressillator (c).
The flat-headed arrows denote negative transcriptional regulation.}
\label{fig:1}
\end{figure}

In this paper we present deterministic analysis (using rate
equations) and stochastic analysis (using direct numerical integration of 
the master equation
and Monte Carlo simulations) of 
three simple genetic circuits:
the autorepressor, the toggle switch and the repressilator
(Fig.~\ref{fig:1}).
We show that fluctuations give rise to both quantitative and
qualitative effect in the dynamics of thse circuits.

The paper is organized as follows.
In Sec. 2 we consider the autorepressor and 
present the deterministic and stochastic methods used in
the paper.
In Sec. 3 we study the genetic switch and
in Sec. 4 we analyze the repressillator circuit.
The results are discussed and summarized in Sec. 5.

%-------------------------------------------------------
\section{The Autorepressor}

The autorepressor is
the simplest genetic circuit that involves feedback.
It consist of a single gene, denoted by $a$, 
that negatively regulates its own expression.
The gene transcribes into mRNAs that translate into $A$ proteins.
These proteins function as repressors. 
When an $A$ protein binds to the $a$ promoter site,
it prevents 
the RNA polymerase from binding
to the promoter and thus inhibits the transcription process.
It turns out that genetic networks include a large
number of autorepressor modules. 
This circuit is thus considered as a network motif
\cite{Rosenfeld2002}.
One may speculate that it performs some crucial function
in the cell.
It was proposed that the role of the autorepressor is to
speed up response times
\cite{Rosenfeld2002} 
or to reduce fluctuations
\cite{Becskei2000}.
Below we analyze the autorepressor using deterministic
and stochastic methods. We utilize its simplicity in order
to present the methodologies in details.  

\subsection{Michaelis-Menten equations}

The dynamics of genetic networks are commonly described
using the Michaelis-Menten equations.
These equations describe the temporal variations in the
concentrations of the relevant molecules in the cell. 
Here, we denote 
the concentration of protein $A$ in a cell
by $[A]$
(by concentration we refer to the average copy number of $A$ proteins
per cell).
The concentration of the corresponding mRNA 
is denoted by $[m]$.
The Michaelis-Menten equations for the concentrations 
take the form

\begin{eqnarray}
\dot{[m]} &=& \frac{g_m}{1+k[A]^n} - d_m [m] \nonumber \\
\dot{[A]} &=& g_p [m] - d_p [A].
\label{eq:michaelis_menten_with_mrna}
\end{eqnarray}

\noindent
The parameters 
$d_m$ 
and
$d_p$ 
(sec$^{-1}$)
are the degradation rates 
of the mRNAs and proteins,
respectively. 
The trascription rate is given by $g_m$
(sec$^{-1}$).
The rate in which each mRNA is translated into proteins
is given by $g_p$
(sec$^{-1}$).
Since $A$ proteins negatively regulate their own synthesis,
the transcription rate is reduced by a factor of 
$1/(1 +k[A]^n)$.
This factor is called the Hill-function.
In this expression, the parameter $k$ 
quantifies the regulation strength 
(determined by the affinity between the repressor and the promoter site).
The exponent $n$ is called the Hill-coefficient.
In general, Hill-function models can be derived from more complete
rate equation models.
In this case,
$n$ is expected to take only
integer values.
In fact, $n$ 
represents the number of copies of the transcription
factor, that are required to be bound simultaneously to the promoter
in order to perform the regulation. 
The case of $n>1$ is often referred to as cooperative binding.
In this paper we consider only integer values of $n$.
However, similar models may also be used to fit empirical data.
In this case, $n$ is simply a fitting parameter which may take 
non-integer values. 

To simplify the analysis of genetic circuits,
the mRNA level is often ignored and
the transcription and translation processes 
are regarded as a single step
of protein synthesis
\cite{Rosenfeld2002}.  
In this case, 
the effective production rate of proteins
is given by 
$g=g_p g_m/d_m$. 
The Michaelis-Menten equations are reduced to

\begin{equation}    
\dot{[A]} = \frac{g}{1+k[A]^n} - d [A],
\label{eq:michaelis_menten_without_mrna}
\end{equation}    

\noindent
where $d=d_p$.
Ignoring the mRNA level is typically 
justified under steady state conditions. 
However, in the analysis of systems 
that are away from steady state 
due to external signals, or those
that exhibit oscillations
or large fluctuations
the mRNA level should be included.
The Michaelis-Menten equations for the
autorepressor exhibit a single steady state solution
for the concentration of $A$ proteins.
In case that $n=1$, it is 
given by

\begin{equation}
\label{eq:repressillator_steady_state_1}
[A] = \frac {-1+\sqrt{1+4kg/d}}{2k}.
\end{equation}

\noindent
This solution is found to be stable 
for any choice of the parameters.

\subsection{Extended set of rate equations}

In the Michaelis-Menten equations the 
negative regulation
process is described by the Hill-function.
This description is rather crude and incomplete.
In order to model 
the regulation process in greater detail 
we present below a more complete rate equation model
\cite{Lipshtat2005}.
In this model, 
we account separably for the populations 
of free and bound proteins.
The bound $A$ proteins are denoted by $r$ and their 
concentration is given by $[r]$.

Consider an autorepressor gene $a$, encoded on the chromosome, which
exhibits no cooperative binding.
In this case the number of bound repressors is in the range
$0 \le [r] \le 1$.
The concentration, $[r]$, can also be considered as 
the fraction of time 
in which the promoter is occupied by a bound repressor
and the transcription process is suppressed. 
Therefore the transcription rate
is reduced by a factor of $(1-[r])$. 
Ignoring the mRNA level,
the extended set of rate equations takes the form
\cite{Lipshtat2005}

\begin{eqnarray}
\dot{[A]} &=& g (1-[r])-d [A]-\alpha_0[A]\left(1-[r]\right)
+\alpha_1[r] \nonumber \\
\dot{[r]} &=& \alpha_0[A]\left(1-[r]\right)-\alpha_1[r],
\label{eq:extended_rate}
\end{eqnarray}

\noindent
where the parameter $\alpha_0$ (sec$^{-1}$) is the binding rate of the 
repressors to the promoter site and $\alpha_1$ (sec$^{-1}$)
is the their unbinding rate.
In the limit in which the binding and unbinding processes 
are much faster than other 
processes in the system 
(namely $\alpha_0,\alpha_1 \gg d,g$),
these equations can be reduced to the Michaelis-Menten form. 
In this limit, the relaxation time of the concentration $[r]$ 
is much shorter than other relaxation times in the system. 
Therefore, one can take the 
time derivative of $[r]$ to zero, 
even if the system is away from steady state.
This brings the rate equations to the 
Michaelis-Menten form 
[Eq. (\ref{eq:michaelis_menten_without_mrna})]
with $n=1$ and 
$k=\alpha_0/\alpha_1$.
Therefore, 
Eqs. (\ref{eq:extended_rate}) 
have the same steady state solution
for the protein $A$ as the Michaelis-Menten equation
(\ref{eq:michaelis_menten_without_mrna}).
However, the dynamics leading to steady state may differ between 
the two equations. 
Furthermore, the extended rate equation model exhibits more
flexibility in the sense that 
it is much easier to insert additional
features into Eqs. (\ref{eq:extended_rate}) than into
Eq. (\ref{eq:michaelis_menten_without_mrna}). 
For example, it is possible to introduce degradation of bound
repressors by adding
the term 
$-d_r [r]$ to the equation for 
$\dot{[r]}$ in
Eqs. (\ref{eq:extended_rate}).

\subsection{Stochastic Analysis}

Transcription factors and other proteins, 
as well as their mRNAs in a cell, 
often appear in low concentrations
\cite{Gillespie1977,McAdams1999,Gibson2001}.
In this case, fluctuations in the copy numbers
of these molecules may play an important role
in the dynamics of genetic networks.
To obtain a better description of thse systems,
one should take into account 
the discrete nature of the molecules
rather than using continuous concentrations.
Moreover, even in case that some transcription  
factor appears in a high concentration,
the regulation is performed by a small number of
copies that are bound to the promoter site.
The fluctuations in the number of bound transcription 
factors give rise to large temporal variations
in the transcription rate of the regulated gene.

In order to account for fluctuations in genetic networks,
stochastic methods are required, such as the master equation
or Monte Carlo simulations
\cite{Arkin1998,Kepler2001,McAdams1997,Paulsson2000,Paulsson2004}.
The master equation is expressed in terms of
the probability distribution 
$P(N_A,N_r)$. 
This is the probability
for a cell to include 
$N_A=0,1,2,\dots$
free copies of protein $A$ 
and $N_r=0,1$ copies of the same protein, 
which are bound to the promoter.
The master equation accounts for the temporal variations in the probability 
distribution. 
For the autorepressor, it takes the form
\cite{Lipshtat2005}

\begin{eqnarray}
\label{eq:autorepressor_master}
\dot P(N_A,N_r) &=& 
 g \delta_{N_r,0} [P(N_A-1,N_r) - P(N_A,N_r)]  \nonumber\\
 &+& d [(N_A+1) P(N_A+1,N_r)      -   N_A  P(N_A,N_r)]       \nonumber\\
 &+& \alpha_0 [\delta_{N_r,1}(N_A+1) P(N_A+1,N_r-1) -  
               \delta_{N_r,0} N_A P(N_A,N_r)]                \nonumber\\
 &+& \alpha_1 [\delta_{N_r,0} P(N_A-1,N_r+1) -
               \delta_{N_r,1} P(N_A,N_r) ],
\end{eqnarray}

\noindent
where the $g$ term accounts for the production of proteins
and the $d$ term accounts for their degradation.
The $\alpha_0$ ($\alpha_1$) terms describe the binding
(unbinding) of proteins to (from) the promoter site.

In numerical integration, the master equation must be truncated
in order to keep the number of equations finite. 
This is done by setting a suitable upper cutoff,
$N_{\rm A}^{\rm max}$, 
on the population size of the free proteins.
In order to maintain the accuracy of the calculations,
the cutoff should be chosen such that
the probability of population sizes beyond it
will be sufficiently small.
The master equation exhibits a single steady state solution,
characterized by 
$\dot{P}(N_A,N_r)=0$
for all $N_A$ and $N_r$.
This solution is always stable 
\cite{VanKampen1992}.
The average concentration of free $A$ 
proteins can be 
obtained from

\begin{equation}
\langle N_{\rm A} \rangle =
\sum_{N_{\rm A}=0}^{N_{\rm A}^{\rm max}} 
\sum_{N_{\rm r}=0}^{1} 
N_{\rm A} P(N_A,N_r).
\label{eq:averagex}
\end{equation}

\noindent
Other properties of the distribution, such as the
variance, can be obtained from the calculation of
higher moments.

Another useful approach to the study of stochastic dynamics 
is provided by Monte Carlo methods
\cite{Gibson2000,Gillespie1977,McAdams1997}.
The simulation dynamics is Markovian. 
At each instant, the
next process to take place is chosen randomly 
from all the possible processes,
where each process is assigned with a suitable weight,
proportional to its rate.
The elapsed time is then updated accordingly.
Unlike the master equation, which accounts for the entire 
distribution, Monte Carlo simulations follow the temporal variations
in protein concentrations in a single cell.

\begin{figure}
\includegraphics[width=13cm]{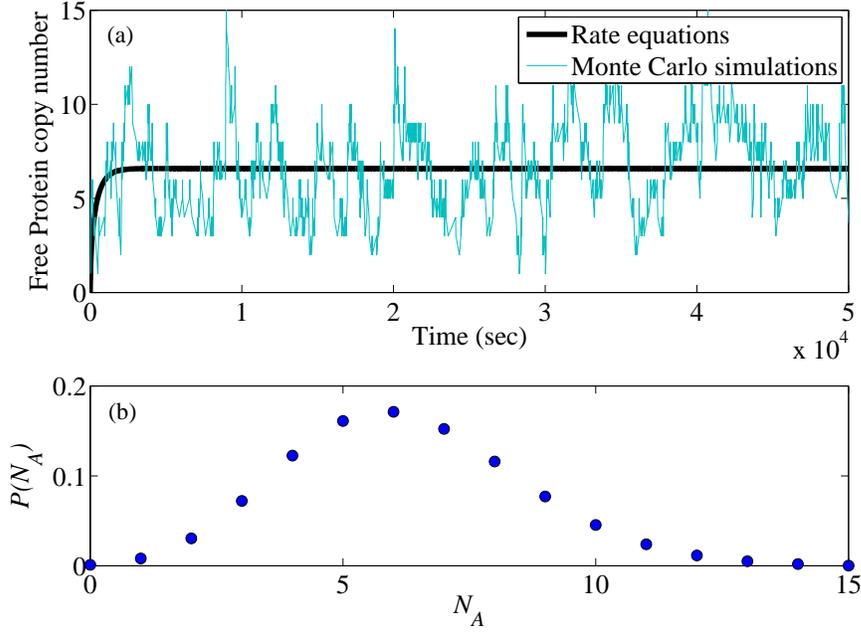}
\caption{
Results for the autorepressor circuit.
(a) The concentration of $A$ proteins vs. time, obtained
from the rate equations and from Monte Carlo simulations.
The rate equation results quickly reach a steady state.
The Monte Carlo results fluctuate around this steady state.
(b) The steady state probability distribution $P(N_A)$
for a cell to contain $N_A$ copies of protein $A$, 
obtained from the master equation.
The parameters used are 
$g=0.05$,
$d=0.001$,
$\alpha_0=0.01$,
$\alpha_1=0.01$
and
$d_r=0$
(sec$^{-1}$).
}
\label{fig:2}
\end{figure}

In 
Fig.~\ref{fig:2}(a)
we present the concentration of $A$ proteins vs. time,
obtained from the rate equations
[Eq. (\ref{eq:extended_rate})]
and from Monte Carlo simulations.
In
Fig.~\ref{fig:2}(b)
we show the probability distribution
$P(N_{\rm A}$
obtained from the master equation
under steady state conditions.

\section{The Genetic Switch}

The genetic toggle switch
consists of two proteins, $A$ and $B$, 
which negatively regulate each other at the transcriptional level
[Fig. \ref{fig:1}(b)].
This architecture may lead to two steady states, 
one dominated by $A$ proteins and the other dominated by $B$ proteins.
When the population of $A$ proteins is much larger than that of $B$ 
proteins, the $A$ proteins tend to suppress the production of $B$ 
proteins. 
Under these conditions, 
the production of $A$ proteins is enhanced, 
because the declining concentration of $B$
proteins is not sufficient to suppress it. 
Therefore, the system approaches a state
rich in $A$ proteins and poor in $B$ proteins. 
Similarly, the system may approach a state rich 
in $B$ proteins and poor in $A$ proteins. 
Transitions between the two states may take place
in response to a suitable external signal.
Spontaneous transitions, due to random fluctuations,
are also possible.
To qualify as a switch, the system should exhibit bistability. 
In the deterministic description, bistability is defined
as the existence of two stable steady state solutions 
of the rate equations.
This description does not account for the possibility of
spontaneous transitions between the two states.
In the stochastic description, spontaneous transitions
are taken into account.
Therefore, the condition for bistability
is that the rate of spontaneous transitions
(due to random fluctuations rather than 
an external signal) is much lower than the rates of
all other relevant processes in the system. 

Genetic switch systems exist in nature, and give
rise to different cell behaviors in different situations.
A notable example is the 
phage $\lambda$ switch
\cite{Ptashne1992}.
This switch appears in $\lambda$
phages, which infect {\it E. coli} bacteria and 
can exist in two exclusive states, one called 
lysogeny and the other called lysis. 
In addition to natural switches, a synthetic switch
was constructed and studied in 
{\it E. coli}
\cite{Gardner2000}.
Numerous studies, 
using rate equations,
have concluded
that cooperative binding is
a necessary condition for the emergence of bistability
\cite{Cherry2000,Gardner2000,Walczak2005,Warren2004,Warren2005}. 
Stochastic analysis reveals the reason to this fact.
For a switch without cooperative binding, three peaks
are obtained in the probability distribution function.
These peaks corresponds to three possible states for the system:
one in which $A$ is highly expressed, 
a second in which $B$ is highly expressed 
and a third in which both proteins are suppressed
(a 'deadlock' situation)
\cite{Lipshtat2006,Loinger2007}. 
Monte Carlo simulations show rapid transitions
between these three states.
The possibility of simultaneous suppression of both proteins, 
prevents the system from functioning as a switch.
It causes the system to exhibit three states 
of limited stability 
instead of the two stable states that are desired.

It is found that in switch systems in which the
$A$ and $B$ repressors exhibit cooperative binding,
the deadlock situation is removed and bistability
emerges.
This can be explained as follows.
The deadlock situation results from a simultaneous 
binding of $A$ and $B$ repressors to the corresponding promoter
sites.
Without cooperative binding, it is sufficient for
the minority protein to recruit a single copy that will bind 
and suppress the production of the dominant protein.
In the case of cooperative binding (for example, with $n=2$)
the minority protein needs to recruit two copies that will
bind simultaneously
in order to suppress the production
of the dominant protein.
This is much less likely. 
Therefore, cooperative binding induces bistability
and enables the system to function as a switch.

Apart from cooperative binding, 
there are several other mechanisms that may stabilize
bistability in genetic switch systems.
The most obvious is the exclusive switch,
where there is an overlap between the promoters 
of $A$ and $B$, leaving no room for both to be 
occupied simultaneously.
Such situations are encountered in nature,  
for example, in the 
lysis-lysogeny switch of phage $\lambda$
\cite{Ptashne1992}.
It was shown that 
in the presence of cooperative binding,
the exclusive binding of $A$ and $B$ repressors
enhances the stability of the
genetic switch
\cite{Warren2004}.
This is because in exclusive binding the access of the
minority specie to the promoter site is blocked by the
dominant specie.
For the exclusive switch without cooperative binding,
rate equations exhibit only a single steady state solution,
namely, the system is not a switch in the deterministic framework.
However, in the stochastic framework the system exhibits 
bistability and functions as a switch.
The distribution $P(N_A,N_B)$ 
exhibits two peaks, one dominated by
$A$ proteins and the other dominated by
$B$ proteins.
The exclusive binding prevents the possibility of a
deadlock situation.

\begin{figure}
\includegraphics[width=13cm]{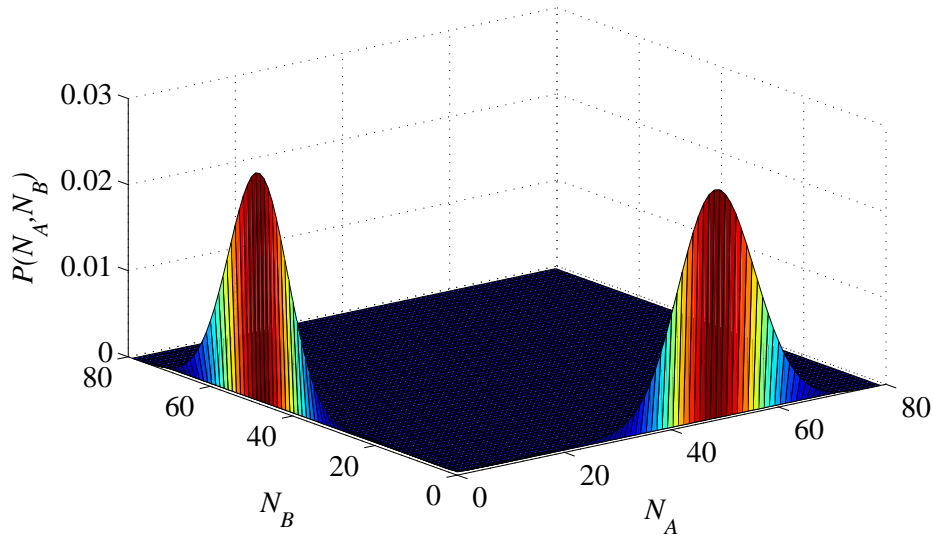}
\caption{
The probability distribution $P(N_{\rm A},N_{\rm B})$
of the concentrations of $A$ and $B$ proteins, for
the switch with degradation of bound repressors,
obtained from the master equation. 
Two sharp peaks are observed, one dominated by $A$ proteins
and the other dominated by $B$ proteins.
The peaks are sharp and are separated by a region
with vanishing probabilities.
As a result, the transition rate between the two
peaks is low and the switch is stable. 
}
\label{fig:3}
\end{figure}

\begin{figure}
\includegraphics[width=13cm]{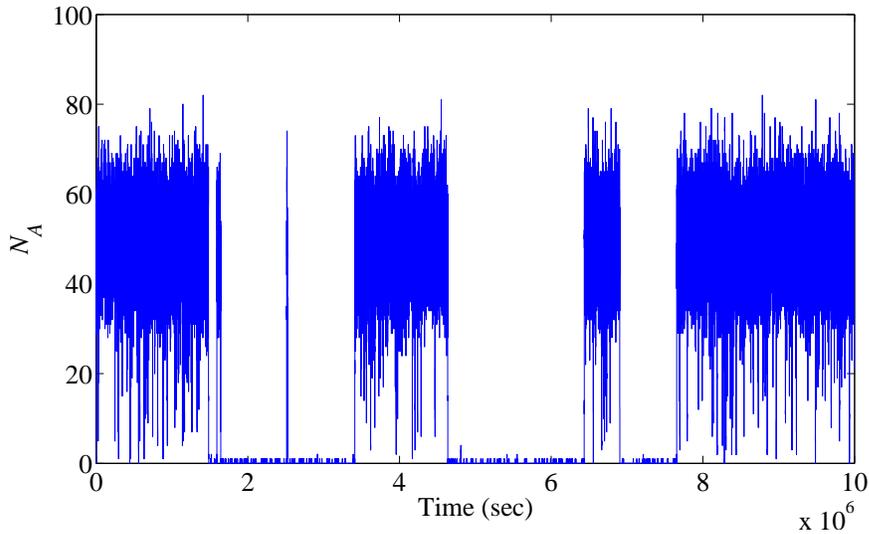}
\caption{
The concentration, $N_A$, of $A$ proteins vs. time,
obtained from Monte Carlo simulations for the switch with
degradation of bound repressors.
The two states are clearly observed: one in which $N_A$  
fluctuates around 50, and another in which it is nearly zero.
Transitions between these states occur 
at an average rate of one transition every 
$\sim 10^6$ sec (about 10 days).
}
\label{fig:4}
\end{figure}

In addition to the exclusive switch, there exist other
variants of the genetic switch circuit 
(we focus here on systems without cooperative binding).
Consider a switch in which not only free proteins,
but also bound proteins experience degradation.
Bound-repressor degradation 
tends to prevent the deadlock situation in which 
both $A$ and $B$ repressors are bound simultaneously. 
This is due to the fact that degradation removes the bound repressor 
from the system, unlike 
unbinding, where the resulting free repressor may quickly bind again.
It turns out that degradation of bound repressors induces 
bistability not only in the stochastic framework 
but also at the level of deterministic rate equations.

In Fig. \ref{fig:3}
we present the 
probability distribution 
$P(N_{\rm A},N_{\rm B})$
of the concentrations of $A$ and $B$ proteins, for
the switch with degradation of bound repressors.
Two sharp peaks, well separated from each other are
observed, illuminating the bistable nature of the
systems.
The parameters used 
in Fig. \ref{fig:3}
and in the rest of the paper
are:
$g=0.15$,
$d=0.003$,
$\alpha_0=0.5$,
$\alpha_1=0.01$
and 
$d_r=0.003$
(sec$^{-1}$).

In Fig. \ref{fig:4}
we present the temporal variations of 
the concentration, 
$N_A$, of $A$ proteins, obtained from Monte Carlo simulations
for the switch with degradation of bound repressors.
Two states are clearly observed: one in which $N_A$ is dominant
and another state in which it is suppressed.
Note that in spite of the very large fluctuations,
the switch is stable and the average time between 
spontaneous transitions is about 10 days.  

Another variant of the genetic switch, 
which exhibits bistability even at the level
of rate equations, 
involves protein-protein interactions,
where $A$ and $B$ proteins bind to each other and
form a complex that does not function as a transcription factor.
This additional process contributes to the stability of the
switch because in such 'pair annihilation' processes
the minority protein is affected more strongly.
It is thus less likely to bind and suppress the production
of the dominant protein.

\section{The Repressilator}

The repressilator circuit consists of three  
transcription factors, 
$A$, $B$ and $C$, which 
negatively regulate each other's synthesis 
in a cyclic manner
[Fig.~\ref{fig:1}(c)].  
This circuit was synthetically constructed 
on plasmids in {\it E. coli} 
and was found to exhibit oscillations
in the concentrations of the three transcription factors. 
To understand the origin of these oscillations, 
consider a situation in which the number of $A$ proteins is large. 
In this case it is likely that one of the $A$ proteins
will bind the to $b$ promoter and will 
repress the production of $B$ proteins.
The reduced level of $B$ proteins will enable the gene
$c$ to be fully expressed and the number of $C$ proteins
will increase and will start to repress gene $a$.
As a result, the number of $A$ proteins will decrease,
and gene $b$ will be activated, completing a full cycle.
The order of appearance of the dominant protein
type in this cycle is $A \to C \to B \to A$. 

\begin{figure}
\includegraphics[width=13cm]{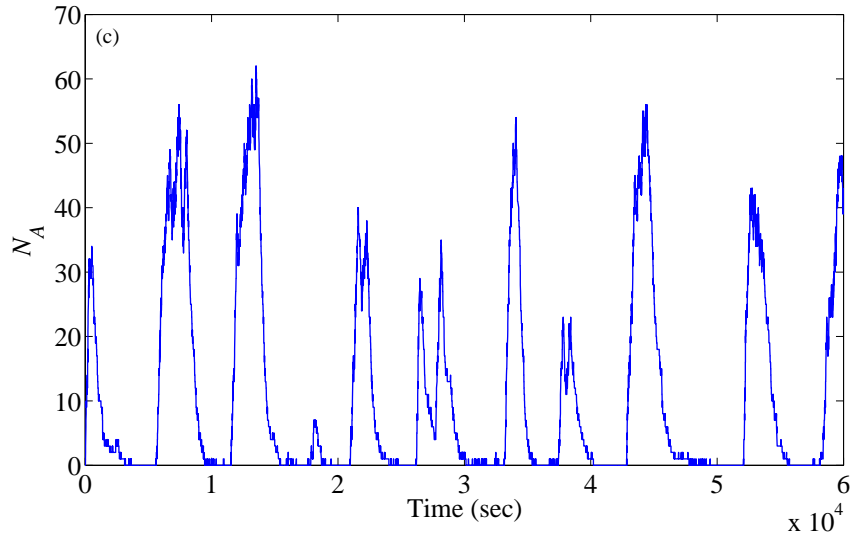}
\caption{
The concentration, $N_A$, of $A$ proteins vs. time, obtained from
Monte Carlo simulations of the repressilator circuit.  
The oscillations are noisy. Their
period and amplitude vary from cycle to cycle.
}
\label{fig:5}
\end{figure}

From a theoretical point of view, oscillations in this
system can be obtained (under some conditions) both in 
rate equations and in Monte Carlo simulations
[Fig. \ref{fig:5}].
The oscillations obtained from the rate equations 
are regular,
forming a stable limit-cycle.
The oscillations
obtained from the Monte Carlo simulations are noisy and irregular.
Moreover, the period and amplitude differ significantly 
between the rate equations and the Monte Carlo simulations
\cite{Loinger2007b}.

The repressilator system was constructed synthetically on
plasmids. When the number of plasmids in a cell is small,
the numbers of promoter sites and bound transcription factors 
are also small.
As a result, one expects large fluctuations in the transcription
rates regulated by these transcription factors. 
In this case stochastic methods are required.
However, when the number of plasmids is large, fluctuations
are reduced and the rate equations become applicable.
Therefore, by gradually changing the number of plasmids in the cell
one can explore the cross-over from the stochastic regime,
where the oscillations are noisy, to the deterministic 
regime, where the oscillations are regular.

\section{Summary}

We have presented deterministic and stochastic analysis of 
three simple genetic circuits which involve transcriptional
regulation and feedback: 
the autorepressor, the switch and the repressilator.
Such systems can be simulated using rate equations, that
account for the concentrations of the mRNAs and proteins produced
by these genes. 
Rate equations 
are suitable when these concentrations are large and 
fluctuations are negligible.
However, when some of the transcription factors and their binding sites
in a cell appear in low copy numbers, 
fluctuations become important
and the rate equations fail. 
In this case stochastic methods such as the master equation or
Monte Carlo simulations are required.
We have shown that fluctuations give rise to quantitative
and qualitative changes in the dynamics of the systems.
In particular, we demonstrated the fluctuations-induced bistability
in the exclusive switch and the noisy oscillations
obtained in the repressilator circuit.

\bibliographystyle{prsty}

\begin{thebibliography}{xx}

\bibitem{Arkin1998}
{A. Arkin, J. Ross and H.H. McAdams}, 
{\it Stochastic kinetic analysis of developmental pathway bifurcation
in $\lambda$-infected E. coli cells}
Genetics {\bf 149} (1998), 1633.

\bibitem{Becskei2000}
{A. Becskei and L. Serrano},   
{\it Engineering stability in gene networks by autoregulation}, 
Nature {\bf 405} (2000), 590.

\bibitem{Cherry2000}
{J.L. Cherry and F.R. Adler}, 
{\it How to make a biological switch}
J. Theor. Biol. {\bf 203} (2000), 117.

\bibitem{Elowitz2000}
{M.B. Elowitz and S. Leibler},    
{\it A synthetic oscillatory network of transcriptional regulators}, 
Nature {\bf 403} (2000), 335.

\bibitem{Elowitz2002}
{M.B. Elowitz, A.J. Levine, E.D. Siggia and P.S. Swain},  
{\it Stochastic gene expression in a single cell}, 
Science {\bf 297} (2002), 1183.

\bibitem{Gardner2000}
{T.S. Gardner, C.R. Cantor and J.J. Collins}, 
{\it Construction of a genetic toggle switch in E. coli}
Nature {\bf 403} (2000), 339.

\bibitem{Gibson2000}
{M.A. Gibson and J. Bruck}, 
{\it Efficient
exact stochastic simulation of chemical systems with many species and many
channels}, 
{J. Phys. Chem.} {\bf 104} (2000), 1876.

\bibitem{Gibson2001}
{M.A. Gibson and E. Mjolsness},
  {\it Modeling the activity of single genes}, {\em in} {J. M. Bower and H.
  Bolouri}, ed., {Computational Modeling of Genetic and Biochemical
  Networks}, {MIT press}, {Cambridge, MA} (2000), pp.~1--48.

\bibitem{Gillespie1977}
{D.T. Gillespie}, 
{\it Exact stochastic simulation of coupled chemical reactions}, 
{\em J. Phys. Chem.} 
{\bf 81} (1977), 2340.

\bibitem{Kepler2001}
{T.B. Kepler and T.C. Elston},
{\it Stochasticity in transcriptional regulation: origins, consequences, and
  mathematical modeling representations}, 
{\em Biophysical Journal} {\bf 81} (2001), 3116.

\bibitem{Laslo2006}
{P. Laslo et al.}, 
{\it Multilineage transcriptional priming and determination of
alternate hematopoetic cell fates},
Cell {\bf 126} (2006), 755.

\bibitem{Lipshtat2005}
{A. Lipshtat, H.B. Perets, N.Q. Balaban and O. Biham}, 
{\it Modeling of negative autoregulated genetic networks in single cells},
Gene {\bf 347} (2005), 265. (See also arxiv:q-bio.MN/0504030)

\bibitem{Lipshtat2006}
{A. Lipshtat, A. Loinger, N.Q. Balaban and O. Biham}, 
{\it Genetic toggle switch without cooperative binding},
Phys. Rev. Lett. {\bf 96} (2006), 188101.

\bibitem{Loinger2007}
{A. Loinger, A. Lipshtat, N.Q. Balaban and O. Biham}, 
{\it Stochastic simulations of genetic switch systems},
Phys. Rev. E {\bf 75} (2007),  021904.

\bibitem{Loinger2007b}
{A. Loinger and O. Biham}, 
{\it Stochastic simulations of the repressilator circuit},
submitted to Phys. Rev. E (2007). 

\bibitem{McAdams1997}
{H.H. McAdams and A. Arkin},
{\it Stochastic mechanisms in gene expression}, 
{\em Proc. Natl. Acad. Sci. US}
  {\bf 94} (1997), 814.

\bibitem{McAdams1999}
{H.H. McAdams and A. Arkin}, {\it It's a
  noisy business! Genetic regulation at the nanomolar scale}, {\em Trends
  Genet.} {\bf 15} (1999), 65.

\bibitem{Milo2002}
{R. Milo, S. Shen-Orr, S. Itzkovitz, N. Kashtan, D. Chklovskii and U. Alon},
{\it Network Motifs: Simple Building
  Blocks of Complex Networks}, {\em Science} {\bf 298} (2002), 824.

\bibitem{Ozbudak2004}
E.M. Ozbudak, M. Thattai, H.N. Lim, B.I. Shraiman and A. van Oudenaarden  
{\it Multistability in the lactose utilization 
network of Escherichia coli} 
{\em Nature} {\bf 427} (2004) 737.

\bibitem{Paulsson2000}
{J. Paulsson and M. Ehrenberg},
{\it Random signal fluctuations can reduce random fluctuations in regulated
components of chemical regulatory networks}, 
{\em Phys. Rev. Lett.} {\bf
  84} (2000), 5447.

\bibitem{Paulsson2000b}
{J. Paulsson}, {\it Stochastic focusing:
  fluctuation-enhanced sensitivity of intracellular regulation}, {\em Proc.
  Natl. Acad. Sci. US} {\bf 97} (2000), 7148.

\bibitem{Paulsson2002}
{J. Paulsson}, {\it Multileveled
selection on plasmid replication}, 
{\em Genetics} {\bf 161} (2002), 1373.

\bibitem{Paulsson2004}
{J. Paulsson}, 
{\it Summing up the noise
in gene networks}, {\em Nature} {\bf 427} (2004), 415.

\bibitem{Ptashne1992}
{M. Ptashne}, 
{\em {A Genetic Switch: Phage $\lambda$ and Higher Organisms, 2nd
  edition.}} ({Cell Press and Blackwell Scientific Publications}, {Cambridge,
  MA}, 1992).

\bibitem{Rosenfeld2002}
{N. Rosenfeld, M.B. Elowitz and U. Alon}, 
{\it Negative autoregulation speeds the response times
of transcription networks}, 
{\em J. Mol. Biol.} {\bf 323} (2002), 785.

\bibitem{Swain2002}
{P.S. Swain, M.B. Elowitz and E.D. Siggia}, 
{\it Intrinsic and extrinsic contributions to
  stochasticity in gene expression}, 
{\em Proc. Natl. Acad. Sci. US} {\bf 99} (2002), 12795.

\bibitem{VanKampen1992}
{N.G. van Kampen}, 
{\em {Stochastic processes in physics and chemistry}}
({North-Holland}, {}, 1992).

\bibitem{Walczak2005}
{A. M. Walczak, M. Sasai, and P. Wolynes}, 
{\it Self consistent proteomic field theory of stochastic gene switches},
Biophysical Journal {\bf 88},  
  (2005) 828.

\bibitem{Warren2004}
{P.B. Warren and P.R. ten Wolde}, 
{\it Enhancement of the stability of genetic switches by overlapping upstream
regulatory domains},
Phys. Rev. Lett. {\bf 92} (2004),  128101.

\bibitem{Warren2005}
{P.B. Warren and P.R. ten Wolde}, 
{\it Chemical models of genetic toggle switches},
J. Phys. Chem. B {\bf 109} (2005), 6812.

\end{thebibliography}

\end{document}